\documentclass[aps,twocolumn]{revtex4}
\usepackage{epsfig}
\newcommand{\be}{\begin{equation}}
\newcommand{\ee}{\end{equation}}
\newcommand{\bea}{\begin{eqnarray}}
\newcommand{\eea}{\end{eqnarray}}
\newcommand{\ba}{\begin{array}}
\newcommand{\ea}{\end{array}}

\begin{document}

\title{Temporal scaling at Feigenbaum points and non-extensive thermodynamics}
\author{Peter Grassberger}
\affiliation{John-von-Neumann Institute for Computing, Forschungszentrum
J\"ulich, D-52425 J\"ulich, Germany\\
Perimeter Institute for Theoretical Physics, Waterloo, Canada, N2L 2Y5} 

\date{\today}
\begin{abstract}
We show that recent claims for the non-stationary behaviour of the 
logistic map at the Feigenbaum point based on non-extensive thermodynamics 
are either wrong or can be 
easily deduced from well-known properties of the Feigenbaum attractor. In 
particular, there is no generalized Pesin identity for this system, the 
existing ``proofs" being based on misconceptions about basic notions of 
ergodic theory. In deriving several new scaling laws of the 
Feigenbaum attractor, thorough use is made of its detailed structure, but 
there is no obvious connection to non-extensive thermodynamics.
\end{abstract}

\maketitle

During the last decade has appeared a vast literature on a new 
``non-extensive thermodynamics" (NET), which uses a maximum entropy 
principle with the Shannon entropy replaced by the Havrda-Charvat 
\cite{Havrda} (``Tsallis") entropy \cite{Tsallis88,Tsallis,Tsallis2004,website}. 
Several versions of NET were proposed by Tsallis and others
\cite{Tsallis}, in order to avoid inconsistencies. But as shown most 
forcefully by Nauenberg \cite{Nauenberg}, none of these versions is 
consistent with thermodynamics, at least for equilibrium systems. 

For non-equilibrium systems the situation is not quite so simple, as 
the standard maximum entropy principle is usually not applicable.
As shown by Jaynes \cite{Jaynes}, the 
formalism of statistical physics can be obtained by Occam's razor: 
If all knowledge can be formulated in terms of constraints, then 
the only rationally justifiable ansatz for the probability distribution 
is the one which maximizes Shannon entropy (which has to be replaced by 
Kullback-Leibler information, if some knowledge existed prior to these
constraints). The main reason why standard 
thermodynamics cannot be applied to most non-equilibrium systems is 
that prior knowledge cannot be cast into the form of a few constraints. 
But there is no reason for abandoning an information theoretic 
interpretation of entropy, and Shannon (Kullback-Leibler) entropy
is the only consistent probabilistic measure of (relative) information.

The main reason why NET is still vigorously pursued is, it seems, the 
claim that it is able to make striking predictions that could not 
be made within a more conventional framework. Typically, these are for
non-equilibrium phenomena with distributions showing power laws with 
heavy tails. In deriving these distributions, not only is Shannon entropy
replaced arbitrarily by Tsallis entropy. Also the constraints are modified in
a way which has no clear rational motivation -- except that one arrives thereby 
at expressions more easily handled.

A careful study of most -- if not all -- examples where NET was supposedly 
successful shows that the success is much less clear than claimed. In a 
later publication, we will substantiate this further by discussing several 
such examples. In the present letter, we 
discuss in depth just one single example, which has been treated
in several papers 
\cite{Tsallis97,Costa,Lyra,Moura,Yang,Buiatti,Montangero,Latora,Tirnakli,Borges,Tsallis2,Baldovin,Ananos,Tonelli2004,Coraddu,Tonelli2005}, 
and which was claimed to show the success of NET in a particularly clear
way.

This example is the non-stationary behaviour of the Feigenbaum attractor
\cite{Feigenbaum}. More precisely, one can study:
(1) sensitive dependence on initial conditions (both for finite and
 infinitesimal perturbations, both on the attractor and in its basin of attraction);
(2) scaling of different dynamical (Shannon, Renyi, Tsallis) entropies of
 various ensembles of trajectories with their time length $T$;
(3) convergence of a typical trajectory (with random initial
condition in its vicinity) to the attractor;
(4) scaling of ``static" (i.e., Boltzmann-Gibbs) entropies of an
ensemble of points with time.


Most of these problems have been discussed since the early 1980's 
\cite{Huberman,Scheunert,Grass86,Anania,Hata,Freund}.
Also, it had been realized from the very beginning that some of these 
questions are subtle due to large fluctuations (``multifractality" of the 
Feigenbaum attractor). In the first papers based on NET 
\cite{Tsallis97,Costa,Lyra} this was missed, leading to wrong claims that 
a single non-extensivity parameter could describe scaling at the onset of 
chaos. Although this was corrected recently, we shall see that 
the effect of fluctuations is still not fully appreciated in some of 
these papers \cite{Ananos,Tonelli2005}.

In the following we give theoretical arguments only for the 
Feigenbaum map \cite{Feigenbaum} $g(x)$ defined by 
$g(g(x)) = -\alpha^{-1} g(\alpha x)$ and $g(x) = 1-cx^2+ O(x^4)$ 
for $x\to 0$, but we use the logistic map $a-x^2$ with 
$a=1.401155189\ldots$ for numerical calculations. Problems of universality 
have been discussed e.g. in \cite{Feigenbaum,Scheunert}.
 
Let us first discuss the dependence on infinitesimal changes in the initial 
conditions, i.e. the behaviour of $\Lambda_n(x_0) = \ln|dx_n/dx_0|$, where 
$x_{i+1} = g(x_i)$. As pointed out in \cite{Scheunert}, $\Lambda_n(x_0)$
fluctuates very strongly with $n$ and $x_0$. Different ways of 
averaging over $x_0$ give therefore rise to different scalings with $n$.
If we take arithmetic averages over $\Lambda_n$ (i.e., geometric averages 
over $|dx_n/dx_0|$), we get
\be
   \int dx_0 w_0(x_0) \ln|dx_n/dx_0| \sim \gamma \ln n         \label{geom}
\ee
with $\gamma = 0.599\pm 0.003$ for all smooth initial distributions $w_0(x)$,
at least when we also do an additional averaging over $n$ to damp out 
remaining oscillations \cite{foot1}. On the other hand, as proven rigorously in 
\cite{Scheunert}, arithmetic averages over $|dx_n/dx_0|$ give
\be
   \int dx_0 w_0(x_0) |dx_n/dx_0| \sim n^{{\rm const} \ln n}.  \label{arith}
\ee

Without averaging over $x_0$ one can of course obtain completely different 
behaviour. For $x_0=1$, e.g., one finds $|dx_n/dx_0| \to const$ for 
$n=2^k, k\to\infty$, while 
\be
   |dx_n/dx_0| = \alpha^k =(n+1)^{\log_2 \alpha}    \label{eq1}
\ee
(exactly) for $n=2^k-1$ \cite{footnote1}. Tsallis {\it et al.} prefer to 
write Eq.(\ref{eq1}) as $|dx_n/dx_0| = (1+\lambda(1-q)n)^{1/(1-q)}$ with
$\lambda = 1/(1-q)=\log_2\alpha$, and call the r.h.s. a $q$-exponential.

\begin{figure}
  \begin{center}
    \psfig{file=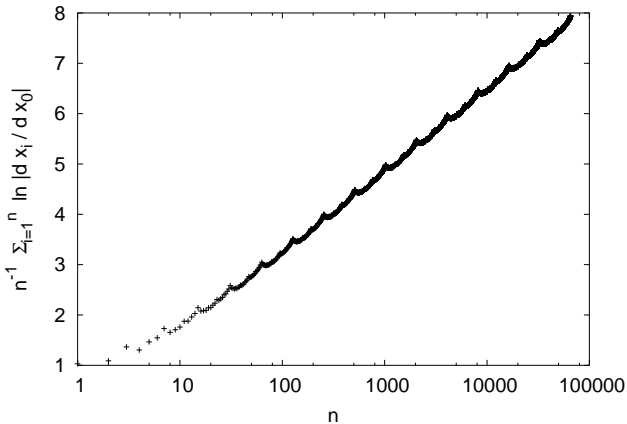,width=5.8cm,angle=270}
   \caption{Scaling of the time averaged local sensitivity exponents, 
    for trajectories starting at $x_0=1$.}
\label{Fig.av-lyapu}
\end{center}
\end{figure}

Notice that Eq.(\ref{eq1}) is {\it not} a scaling law, since it holds only for 
special values of $n$. But one obtains a scaling law by taking averages
over $n$ (see Fig.~1),
\be
   \bar{\Lambda}_n(x_0) \equiv n^{-1} \sum_{i=1}^n \Lambda_i(x_0) \sim \beta \ln n \quad {\rm for}\;\; x_0=1
                    \label{eq1a}
\ee
with $\beta = 0.71339380\ldots$. Notice that the constants in Eqs.(\ref{geom}), 
(\ref{eq1}), and (\ref{eq1a}) are not directly related.

Defining $\bar{\Lambda}_n(x)$ for any $x$ on the attractor 
as in Eq(\ref{eq1a}) gives the natural generalization of the Lyapunov exponent. It fluctuates 
strongly with $x$ \cite{Anania,foot1} and for 
many applications like Pesin's identity \cite{note3} one needs averages over $x$ 
with respect to the natural measure $\mu(x)$. In the present case, one can easily see that
the average
\be
   \langle \bar{\Lambda}_n\rangle \equiv \int d\mu(x) \bar{\Lambda}_n(x) 
\ee
is identically zero for all $n$ \cite{footnote2}, which indicates that there cannot 
exist a close analogy to Pesin's identity. The identities suggested and verified
in \cite{Tsallis97,Baldovin04,Ananos} are trivialities where the authors started with 
an initial distribution $w_0(x)$ narrowly localized around $x_0$ and followed the 
evolution only up to values of $n$ for which $w_n(x)$ is still smooth and described 
by the flow linearized around the trajectory starting at $x_0$ \cite{note-KS}. 
In addition they considered, 
instead of the Kolmogorov-Sinai (KS) entropy, just the difference $S_n-S_0$ between two 
static ``Boltzmann-Gibbs" entropies 
\be
   S_n = -\int dx\; w_n(x) \ln w_n(x).             \label{BG}
\ee
In this case one has of course $S_n-S_0 = \Lambda_n(x_0)$, but this has no connection 
to any (generalized) Pesin identity. 

Eqs.(\ref{eq1}) and (\ref{eq1a}) apply to trajectories starting 
{\it on the attractor}. In order to understand the origin of 
Eqs.(\ref{geom}) and (\ref{arith}) 
one has to study how trajectories starting in its vicinity approach the attractor. 
For this one has to use the detailed triadic Cantor structure of the Feigenbaum
attractor \cite{Feigenbaum}.
Associated to this structure is a set of open disjoint intervals $I_{k,i}$, 
the closure of which covers the interval $[-1/\alpha,1]$ which contains the 
attractor \cite{Scheunert}. $I(0,1)$ is the hole cut out from the middle of 
the Cantor set in the first stage of construction; $I(1,1)$ and $I(1,2)$ are 
the second generation holes, $I(2,1)$ to $I(2,4)$ are the holes cut out in the 
third step, etc. Each $I_{k,i}$ contains exactly one point on the instable 
periodic orbit of period $2^k$. 
Let us define $I_k = \bigcup_i I_{k,i}$, i.e. $I_k$ is all what 
is cut out during the $k-$th step. The evolution of a typical trajectory can 
then be viewed as a tumbling through the $I_k$'s, with $k$ never decreasing
with $n$. The average increase of $k$ is for large $n$ exactly given by 
\be
   \langle k\rangle = {\rm const} + \log_2 n.      \label{kav}
\ee

As shown in \cite{Scheunert}, Eq.(\ref{geom}) follows (up to the precise value 
of $\gamma$) from Eq.(\ref{kav}) and from the scaling behaviour proven by 
Feigenbaum. 

In \cite{Ananos} the authors studied another average over $\Lambda_n(x)$, 
in between Eqs.(\ref{geom}) and (\ref{arith}), 
\be
   \xi_q(n) \equiv \int dx_0 w_0(x_0) |dx_n/dx_0|^{1-q}.    \label{xi}
\ee
From straightforward simulations they concluded that $\xi_q(n)$
becomes (asymptotically) linear in $n$, $\xi_q(n) \sim n$, for $q=0.36$ and 
called this value $q_{\rm sen}^{\rm av}$. But neither analytic nor numerical
estimates of $\xi_q(n)$ seem easy. A direct numerical estimate
as in \cite{Ananos} is prone to large errors, since the integrand of
Eq.(\ref{xi}) (with constant $w_0(x_0)$ and for $q=q_{\rm sen}^{\rm av}$)
is very sharply peaked near the unstable periodic orbits of small periods. 
Choosing initial values $x_0$ at random one will miss these peaks, unless 
one has extremely high statistics.
But it is not clear anyhow why Eq.(\ref{xi}) should be of interest. In 
contrast to what its name suggests, $q_{\rm sen}^{\rm av}$ does not measure 
{\it the} average sensitivity to initial conditions but just one particular
average.

Another consequence of Eq.(\ref{kav}) is that the (geometrically) average 
distance from the attractor of a point starting randomly in $[-1/\alpha,1]$ 
decreases as \cite{Scheunert}
\be
   d_n \sim n^{-1/D_1}
\ee
where $D_1=0.517097\ldots$ is the information dimension of the Feigenbaum 
attractor.

\begin{figure}
  \begin{center}
    \psfig{file=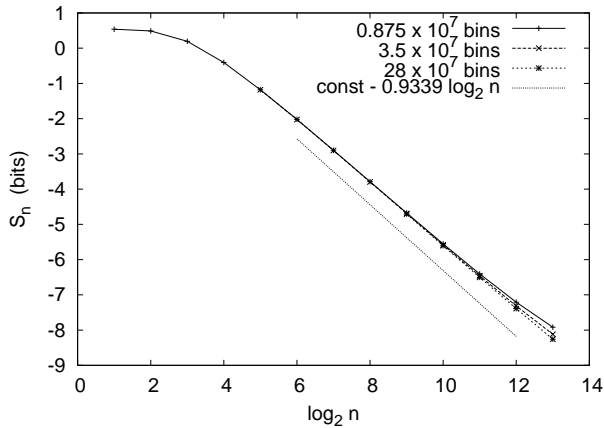,width=5.8cm,angle=270}
   \caption{Time dependence of the ``Boltzmann-Gibbs" entropy $S_n$, i.e. of 
    the differential Shannon entropy of the $x$-distribution, starting with a 
    uniform distribution over the interval $[a-a^2,a]$. The three curves
    correspond to different binnings. The lines connecting the data points
    are just to guide the eye, and omit the lognormal oscillations. 
    The straight line indicates the theoretically predicted slope.} 
\label{Fig.Sn}
\end{center}
\end{figure}
\vglue -1pt

The same argument can also be immediately used to derive the scaling of the 
``Boltzmann-Gibbs" entropy with $n$, after starting with a random ensemble
\cite{Tonelli2005}. Assume we have $N_n$ points distributed randomly with 
respect to $w_n(x)$. The distance between point $x_n^{(j)}$ and its nearest 
neighbour is called $r_j$. The Kozachenko-Leonenko estimator \cite{Koza} for 
differential entropies gives then for large $N_n$
\be
   S_n = {\rm const} + \ln N_n + \sum_{i=j}^{N_n} \ln r_j\;.
\ee
To apply this for the present problem, we choose
furthermore $N_n =n N_0$ so that $N_n$ scales as the number of intervals $I_{k,i}$
with $k=const+\log_2 n$. We notice also that all $I_{k,i}$ with $i=1,\ldots 2^k$ 
have roughly the same statistical weight. Then the distance to the nearest 
neighbour scales as the size of the interval in which the point is located, 
$ r_j \in |I_{k,i}|$  with $ x_j \in I_{k,i}$, and it follows that
\be
   S_n = {\rm const} - {1-D_1\over D_1} \log_2 n.     \label{Sn}
\ee
Data obtained by spraying $N$ points uniformly onto $[a-a^2,a]$, letting them
evolve according to the critical logistic map, and estimating $S_n$ by binning
this interval into $M$ bins, are shown in Fig.~\ref{Fig.Sn}. Here we used $N>2\times 10^9$ 
which is largely sufficient for convergence (we used the entropy estimator 
corrections given in \cite{grass-entropy}). Convergence with the number of 
bins is much slower. We see very clear changes as we increase $M$ 
from $\approx 10^7$ to $\approx 3\times 10^8$. This explains also the 
small remaining discrepancy.

Although $S_n$ had been discussed as an interesting quantity in the NET literature 
\cite{Moura,Borges,Coraddu,Tonelli2005}, we are not aware of a previous estimate. 
Instead, several authors \cite{Moura, Borges,Tonelli2005} have studied a different 
quantity $W_n$, which they supposed to have the same scaling. Instead of taking a 
sum over all non-empty bins with weights $p_i\log(1/p_i)$, $W_n$ is just the number
of non-empty bins. Scaling of this quantity is much more subtle -- both 
theoretically and numerically. We should expect larger finite-$n$, finite $M$, and 
finite $N$ corrections for $W_n$ than for $S_n$, but when $n, M$, and $N$ are 
sufficiently large \cite{note5}, we can apply basically the same reasoning to $W_n$ 
as we did for $S_n$: It scales like the total length of all intervals $I_{k,i}$
with $k = \log_2 n$. Calculating the latter is easy and gives
\be
   W_n \sim n^{-0.800138194\ldots}\,.      \label{Wn}
\ee
The exponent measured in \cite{Moura} was quite different ($0.71\pm 0.01$ instead of 0.80),
but this is easily explained by the expected large corrections to scaling, which
have even led to claims of non-universality in \cite{Tonelli2005}.

The KS entropy for a one-humped map with maximum at $x=0$ 
is obtained from the sequences $s_n = ({\rm sign}(x_1),\ldots {\rm sign}(x_n))$, as 
\be
   H_{KS} = \lim_{n\to\infty} n^{-1} H_n
\ee
with $H_n = -\sum_{s_n} p(s_n) \log  p(s_n)$. As shown in \cite{Grass86,Freund},
for the Feigenbaum attractor one has $H_n = \log n + O(1)$. Thus the KS
entropy is zero, but there is a logarithmic increase of $H_n$. With the same 
methods one can prove {\it exactly}, that all Renyi entropies $H^q = 
\lim_{n\to\infty}H_n^q$ with $H_n^q = (1-q)^{-1}\log \sum_{s_n} [p(s_n)]^q$ 
are equal for this problem, $\lim_{n\to\infty} H_n^q /H_n =1$ 
for all $q$ \cite{note_Hn}. For Tsallis-type generalized 
KS entropies, this gives $K_n^q \equiv (1-q)^{-1}(\exp((1-q)H_n^q)-1)\sim 
n^{1-q}$, at variance with \cite{Montangero}.

The $q$-independence of the Renyi entropies is surprising in view of 
the multifractality of the attractor. For a chaotic attractor, the $H_n^q$ are 
closely related, via generalizations of Pesin's identity, to $q$-th moments of 
$\Lambda(x)$ \cite{Badii}.  This shows again that the onset of chaos is more 
subtle than expected in the NET literature.

In the first papers \cite{Tsallis97,Costa,Lyra,Moura}, it was supposed but 
never substantiated that the parameter $q$ of NET can be obtained, also 
for the Feigenbaum map, by some maximum entropy principle. 
Although this was never withdrawn, $q$ is now fixed such that one 
obtains linear time dependencies, when logarithms 
are replaced by $q-$deformed logarithms (see e.g. \cite{Ananos}). The reason
for this is not clear, since there is no need for all time dependencies to 
be linear. But even worse, with
the proliferation of different (although closely related) scaling laws, one
obtains -- for the single case of the Feigenbaum map! -- a rich zoo of different 
$q$ values \cite{Ananos,Tsallis2004}. With the additional scaling laws found in
the present paper, there would be even more $q$'s -- unless one accepts at last 
that non-extensive thermodynamics is (at least at the onset of chaos) just 
a chimera. On the other hand, we 
consider these new scaling laws as an important part of the present letter.

One reason for preferring ``$q$-exponentials" (i.e., generalized Pareto 
distributions \cite{Pickands}) over power laws could be that the former
give correct deviations from pure power laws at small arguments. But in all cases
studied in the present paper (except Eq.(\ref{eq1})) the small-$n$ limits are 
neither given by pure powers nor by $q$-exponentials.

Finally, as a last remark: In \cite{Latora,Tsallis,Ananos} the Feigenbaum map 
was chosen as the prototype of a supposedly {\it weakly mixing} system. But 
it is not mixing at all.

I am indebted to Maya Paczuski, Walter Nadler and Karol Zyczkowski for numerous 
discussions.


\begin{thebibliography}{}
\bibitem{foot1} These oscillations and the ones seen in Fig.~1 are due to the Cantor structure
   of the Feigenbaum attractor and will not be discussed further.
\bibitem{footnote1} Eq.(\ref{eq1}) follows immediately from the 
   Feigenbaum-Cvitanovic relation $\alpha g(g(x)) = -g(\alpha x)$ 
   and from the well known facts \cite{Feigenbaum} that $g(1)=-1/\alpha, g'(1)=-\alpha$,
   and $g'(-1/\alpha)=1$.
   The equation for $n=2^k$ uses in addition the fact that $x_{(2^k-1)}=g^{(2^k-1)}(1)$
   is close to zero, where $g(x)$ has its quadratic maximum. See also \cite{Baldovin}. 
   For Feigenbaum type maps $g_z(x)$ with maxima of order $z$, the same easy derivation 
   gives $|dx_n/dx_0| = \alpha^{(z-1)k}$. For no obvious reasons, this relation was claimed 
   to show the importance of NET in \cite{Lyra}.
\bibitem{footnote2} The natural measure can be approximated as $d\mu(x)/dx \approx
   N^{-1} \sum_{i=N+1}^{2N} \delta(x-g^{(i)}(1))$ for $N>>1$. Thus
   $\langle \bar{\Lambda}_n \rangle$ is given by the increase of the curve shown in 
   Fig.~1 when the x-coordinate increases by $n$, averaged over 
   $[N+1,2N]$ with $N>>n$, and then divided by $n$. In the limit $n/N\to 0$
   this obviously converges to zero.
\bibitem{note5} In contrast to Eq.(\ref{Sn}), where one has to 
   take first $N\to\infty$, then $M\to\infty$, and at last $n\to\infty$, these
   limits have to be taken simultaneously in Eq.(\ref{Wn}). 
   Taking e.g. $N\to\infty$ first, all $W_n$ would be equal to $W_0$. Eq.(\ref{Wn})
   is obtained with $M> n^{\log_2\alpha}$ and taking $N/M\to \infty$ after 
   $n\to\infty$. It relies on the fact that the distribution of $k$-values 
   for fixed $n$ is extremely sharply cut off at small $k$ \cite{Scheunert}.
\bibitem{note_Hn} If we start on the attractor, and if $n=2^k\; (k=1,2,\ldots)$, 
   then there are $3k$ equally probable $s_n$, giving $H_n = \log (3n/2)$.
   For other values of $n$ there two classes of allowed sequences $s_n$, with 
   the sequences in the first class twice as probable as those in the second class.
\end{thebibliography}
\end{document}